\documentclass[useAMS,usenatbib]{mn2e}
\usepackage{graphicx}
\usepackage{url}
\usepackage{epstopdf}
\usepackage{color}
\usepackage{bm}
\usepackage{amssymb}
\usepackage{amsmath}
\usepackage{hyperref}
\usepackage{textcomp}
\usepackage{natbib}
\usepackage{color}
\usepackage{epsfig}
\usepackage{amsmath}

\hypersetup{
    unicode=true,          
    linktocpage=true,
    pdftoolbar=true,        
   pdfmenubar=true,        
    pdffitwindow=true,     
    pdfstartview={FitH},    
    pdfhighlight={/I},
   hyperindex=true,
    pdftitle={Probing the bias of radio sources at high redshift},    
   pdfauthor={Sean Passmoor},     
   pdfsubject={Probing the bias of radio sources at high redshift},   
   pdfcreator={Sean Passmoor},   
    pdfproducer={Producer}, 
    pdfkeywords={keywords}, 
    pdfnewwindow=true,      
   colorlinks=true,       
    linkcolor=red,          
    citecolor=blue,        
    filecolor=magenta,      
    urlcolor=cyan           
}

\newcommand{\apj}{Astrophys.~J.}

\newcommand{\apjs}{ApJ Suppl.}

\newcommand{\mnras}{MNRAS}

\newcommand{\aap}{A\&A}
\newcommand{\aaps}{A\&AS}
\newcommand{\aj}{A J}

\newcommand{\etal}{{\em et al. }}
\newcommand{\jcap}{J. Cosmology Astropart. Phys.}

\newcommand{\scubed}{S$^3$}
\newcommand{\AF}[1]{#1}
\newcommand{\MS}{}

\newcommand{\SP}[1]{#1}
\newcommand{\CCnew}[1]{#1}
\newcommand{\SPnew}[1]{#1}
\newcommand{\seannew}[1]{#1}
\newcommand{\CC}[1]{#1}
\title[Probing the bias of radio sources at high redshift]{Probing the bias of radio sources at high redshift}

 \author[Passmoor, S. \it{et al.}]{{Sean Passmoor$^{1,2,3}$\thanks{sean@ska.ac.za},Catherine Cress$^{2,3}$, Andreas Faltenbacher$^2$,Russell Johnston$^2$,}
 \newauthor{Mathew Smith$^{2,4}$, Ando Ratsimbazafy$^2$ and Ben Hoyle$^5$}
\\
$^1$SKA SA , The Park, Park Road, Pinelands  7405, South Africa\\
$^2$Department of Physics, University of Western Cape, Bellville 7530, Cape Town, South Africa\\
$^3$Centre for High Performance Computing, 15 Lower Hope Street, Cape Town 7700, South Africa\\
$^4$Astrophysics, Cosmology and Gravity Centre, Department of Mathematics, University of Cape Town, Rondebosch, 7700, South Africa\\
$^5$Institute for Sciences of the Cosmos (ICCUB), University of Barcelona, Marti i Franques 1, Barcelona, 08024 Spain
}
\begin{document}
\date{\today}
\pagerange{\pageref{firstpage}--\pageref{lastpage}} \pubyear{2012}
\label{firstpage}
\maketitle
\begin{abstract}
\CC {The relationship between the clustering of dark matter and that of luminous matter is often described using the bias parameter. Here, } we provide a new method to probe the bias 
of intermediate to high-redshift radio continuum sources for which no redshift information is available. 
We matched radio sources from  the  Faint Images of the Radio Sky at Twenty 
centimetres (FIRST) survey data to their optical counterparts 
in the Sloan Digital Sky Survey (SDSS) to obtain \MS{photometric} redshifts for the \MS{matched} radio sources.
We then use the \AF{publicly available semi-empirical simulation of extragalactic radio continuum sources (\scubed)}
to infer  the  redshift distribution for all FIRST sources and 
estimate the redshift distribution of unmatched sources by subtracting the matched distribution from the distribution of all sources. 
We infer that the majority of unmatched sources are  at higher redshifts than the optically matched sources and demonstrate 
how the angular scales of the angular two-point correlation function can be used to probe different redshift
 ranges. \CC{We compare the angular clustering of radio sources with that expected for dark matter and estimate the bias of different samples.}
\end{abstract}
\section{Introduction}

\MS{Current and future} radio continuum surveys typically probe redshifts out to $z\sim5$ and
often cover a significant fraction of the sky. The large volumes
accessible in these surveys provide a probe of the
large-scale structure and thus can be utilised to test \MS{cosmological models}. One
of the most common approaches to investigate the large-scale
distribution of cosmological objects is the two-point angular
correlation function (ACF) which quantifies the projected clustering of galaxies
on the plane of
the sky. To gain information on the three dimensional distribution \MS{of galaxies} and
their evolution with time, the redshift distribution of the sample needs
to be known. However, in general, redshifts can not be obtained from
radio continuum surveys since 
\AF{the spectra do not show emission or absorption line
  features}.
One way to gain redshift information of
these radio sources is to match them to their optical counterparts for which
the redshifts are known.

First attempts to detect clustering in radio surveys were carried out
in the 1970s, but it was only in 1996 \citep{Cress-96} that the first
high-significance detection of the clustering was made using the Faint
Images of the Radio Sky at Twenty centimetres (FIRST) survey
\citep{Becker-White-Helfand-94}. They found that on angular scales
that probe large-scale structure, the ACF of
galaxies detected down to 1~mJy at 1.4 GHz is well-represented by a
power-law, with a slope somewhat steeper than that found for typical
optical surveys. A number of other studies, e.g \cite{Overzier-03}
and \cite{Blake-Wall-02} also measured clustering of radio sources
using the ACF in the FIRST survey,
in the NRAO VLA Sky Survey (NVSS, \citealt{Condon-98}) and in the
Westerbork Northern Sky Survey (WENSS, \citealt{Rengelink-97}).
Whilst there was some disagreement about the slope of the correlation
function on larger  angular scales, later work by
\cite{Blake-Mauch-Sadler-04} highlighted problems with their earlier
results (associated with over-cleaning of potential sidelobe sources)
and obtained results from all the surveys consistent with
\cite{Cress-96}.

In essence,  all these studies are confined to the investigation of the
projected clustering signal, since many of the sources are too faint in
the optical/IR to \MS{obtain accurate redshifts}. However, some information on
real-space clustering can be inferred, but this relies on estimates of
the average redshift distributions of the sources.

During the 1990s, \cite{Dunlop-Peacock-90} developed models to infer
the redshift distribution of faint radio sources extrapolating from
data at much higher flux densities. Since then, a number of
observations have improved our knowledge in this area.
\citet{Waddington-01} estimated redshifts of a complete sample of 72
radio galaxies down to 1 mJy in about one square degree (65\% with
spectroscopic redshifts). In the Combined EIS-NVSS Survey Of Radio
Sources (CENSORS, \citealt{Best-03,Brookes-06,Brookes-08}), redshifts
were estimated for 150 sources, in a 6 square degree region, with flux
densities above 7.2 mJy in NVSS (63\% of them secure spectroscopic
redshifts). \citet{Magliocchetti-04} studied the optical matches of
FIRST sources in the 2dF survey \citep{Colless-99} and
\citet{Mauch-Sadler-07} studied NVSS matches with $K<12.75$~mag in the
6dF survey \citep{Wakamatsu-03}. These studies all confirmed
the picture
that mJy-radio surveys contain a heterogenous population of galaxies
that is dominated by AGN at higher flux densities and includes
significant fractions of fainter star-forming galaxies at lower
redshifts. They also appeared to rule out a large `spike' of very low-z objects
predicted by some of the Dunlop and Peacock models.

Understanding the nature of the sources in the radio surveys
contributes to our knowledge of the $bias$ of the sources i.e. the
clustering strength of the sources relative to clustering strength of
the underlying dark matter. Knowing the bias is essential for using
clustering as a cosmological probe as it enters into measurements of
autocorrelations, the Integrated-Sachs Wolf (ISW) effect and the lensing effect. However,
little is known about the bias of radio sources.
\citet{Cress-Kamionkowski-98} presented estimates of the bias based on
the FIRST sources. Since then, different and  sometimes contradictory
prescriptions for the bias of radio sources have been used
\citep[e.g.,][]{Raccanelli-08,Raccanelli-11}.
\cite{Wilman-10} utilised a semi-empirical approach
with a bias prescription based on the work of \cite{Mo-White-96}
to predict the clustering of radio sources in future radio surveys.
\SP{The bias value in these models is artificially kept from rising to ``non-physical'' levels which underscores the lack of understanding of the bias of radio sources.}

Future radio surveys  carried out by the Square  Kilometre Array\footnote{http://www.skatelescope.org} (SKA)
will  potentially reach  1  nJy, providing  catalogs  of sources  over
3$\pi$ of  the sky.  SKA Pathfinders such  as the LOw  Frequency ARray\footnote{http://www.lofar.org}
(LOFAR), the Australian Square Kilometre Array Pathfinder (ASKAP), the
South  African Karoo  Array  Telescope (MeerKAT),  the  Westerbork
Synthesis Radio Telescope (WSRT)  using the Apertif instrument and the
extended  Very  Large Array  (eVLA)  will  soon  provide surveys  with
unprecedented   depth  and/or   sensitivity.    The  resulting   radio
auto-correlations and  cross-correlations with other datasets  such as the
CMB can provide valuable tests of cosmology.  They can shed light on the
question   of  non-gaussian   initial  conditions   in   the  universe
\citep{Xia-10}  and   on  issues   concerning  Dark  Energy   via  the
ISW  effect
\citep[e.g.][]{Nolta-04,Raccanelli-08}. They  may also provide strong tests
of modified gravity \citep[e.g.][]{Raccanelli-11} and  be used as
direct  probe of  dark  matter through  gravitational lensing  effects
\citep[e.g.][]{Carilli-Rawlings-04,Kamionkowski-98,Raccanelli-11}.
It is essential for these studies  to have a good understanding of the
underlying bias \MS{of radio galaxies}. \CC{In recent studies, \citep[e.g.][]{Raccanelli-11}, predictions for future constraints on cosmology have been made by marginalizing over a single bias parameter but this does not capture the uncertainties in the evolution of bias which could be very important for the interpretation of measurements.}

Therefore, in this article we  attempt to make a direct  measurement
of the bias of FIRST radio sources  at intermediate redshifts. We match FIRST
sources to galaxies in the Sloan Digital Sky Survey Data Release 7 
(SDSS-DR7, e.g. \citealt{Abazajian:2009ApJS..182..543A} )and determine the
redshift distribution of the  matched sources.   We then  create a
catalog  of  unmatched  sources to probe the higher-z population.

The format  of the paper is as follows. In \S~\ref{sec:data} we
discuss the data and our methodology;  in \S~\ref{sec:results} we discuss the results and present
an estimate of the bias of radio sources at high redshift.  Finally, in \S~\ref{sec:con}
present our conclusions.

\section{Data and methodology}
\label{sec:data}
Our approach to isolating a high-z sample of FIRST sources and
estimating its redshift distribution can be summarised in the
following steps: 
\begin{enumerate}
\item Match the  FIRST sources to galaxies from the SDSS survey   and establish
the redshift distribution of the matches from an SDSS photometric redshift catalogue
\item Use the {\scubed} simulations \citep{Wilman-10} to estimate  an  average  redshift distribution for all FIRST sources.
\item Estimate the redshift distribution of unmatched sources by 
removing the matched distribution from the distribution of all sources. 
 It is then inferred that the  unmatched sources are mostly at higher redshifts.
\item The angular clustering of the high-z sample can then be measured and compared with what is expected
for Dark Matter sampling the same redshift range, to obtain an estimate of the bias.
\end{enumerate}
\subsection{Creating the catalogues}

\subsubsection{The FIRST survey selection}
\label{sec:FIRST_survey_sel}
\AF{In this section we describe the sample selection of the radio
  sources. Table~\ref{table:collapsing} summarises our selection criteria quoted
  below.} The FIRST survey mapped a region of the sky covering
10,000~${\rm deg}^2$ in the Northern Galactic Cap at 1.4 GHz down to
1.4 mJy. The final catalogue contained a total of 816,331~sources \SPnew{with a  completeness of 95\%  down the lower flux level used of 2 mJy}. 

Creating our sample of FIRST sources to be matched to SDSS required various
steps to minimise potential sources of contamination. In the first step, we
removed objects with a high probability of being a sidelobe. The FIRST
survey has assigned to each source a probability of being a sidelobe ranging
from 0 (indicating an object is not a sidelobe) to 1.0 (indicating an object
is a sidelobe). To reduce this source of contamination, we explored various
sidelobe probability values on our initial clustering analysis. This is
discussed in more detail in \S~\ref{sec:results}. However, we note that for our
final selection we found a sidelobe probability value of 0.7 led to \SPnew{results the had minimal effects from the presents for sidelobes or the over-cleaning of them} .
For a sidelobe probability of 0.7 we were left with 795,453 sources.

The next step required the collapsing of multiple components (e.g. double lobes)
 to a single source. Following \cite{Cress-96} we chose a
collapsing radius of 72". \SPnew{This is the linking length of the friends-of-friends algorithm the we use to generate the groups of sources.}
We found that the average collapsed group had 2 to 3 components and a few
groups that had up to 20 components.
To compute the flux for each collapsed source, the integrated flux of each
component was added together. The flux-weighted average positions were then
calculated and used to match with the SDSS. This collapsing radius reduced the
sample to 647,985 sources.

\begin{table}\label{table:collapsing}
\footnotesize
  \begin{center}
    \tabcolsep 2pt
  \caption[Table detailing \MS{the number of sources that satisfy our} source collapsing, area selection and minimum flux cuts]{Detailing \MS{the number of sources that satisfy our}
source collapsing, area selection and minimum flux cuts.}
\label{table:collapsing}
      \begin{tabular}{llcll}
      \\
     	 \hline
           \hline
      ~~~Radio sample &~~~~~Numbers~~~~
      \\
      	\hline
      	\hline
\\
~~~Total FIRST						&~~~~816,331\\
~~~{\bf No. of sources after side-lobe removal}&~~~~{\bf795,453}\\
~~~Collapsing sources in groups $<$ 72"	&~~~~253,971\\
~~~~ collapsed sources					&~~~~106,503\\
~~~~ single sources 					&~~~~541,482\\
~~~{\bf No. of sources after collapsing} 	&~~~~{\bf647,985}
\\
\\
~~~ {\bf No. of sources in selected area }\\
~~~{\bf (130 $\le$ RA $\le$ 240,  5 $\le$ Dec $\le$ 55)} &~~~~{\bf307,859}\\
\\
~~~{\bf No. of sources $\ge$ 2 mJy} 	&~~~~{\bf219,060}\\
      \hline
    \end{tabular}
  \end{center}
\end{table}
Furthermore we only take into account sources within a region
that avoided both gaps in data and the edges of the SDSS and FIRST
surveys. This region is defined by, 130 $\le$ RA $\le$ 240, 5 $\le$
Dec $\le$ 55, covering a total area of 4613.43~${\rm deg}^2$. Our
final catalogue of FIRST sources to be matched with SDSS contained a
total of 307,859 objects. 

Finally, \CCnew{in an attempt to minimize}  effects due to fluctuations in
  sensitivity noted in \citealt{Blake-Mauch-Sadler-04} 
we applied a 2~mJy flux cut which is more than $10$ times the RMS
fluctuations in the considered region. This leaves us with 219,060
sources.

In an attempt to isolate the AGN in the sample and exclude most of the
low-z star-forming galaxies, we consider a sample containing only
sources with flux densities greater than 7~mJy \cite{Waddington-01} \SPnew{this also allows us to compare the redshift distribution to the CENSORS survey}. 
This leaves us with 93,202 sources in the 7~mJy subsample.

\begin{figure}
 \begin{center}
 \includegraphics[width=0.49\textwidth]{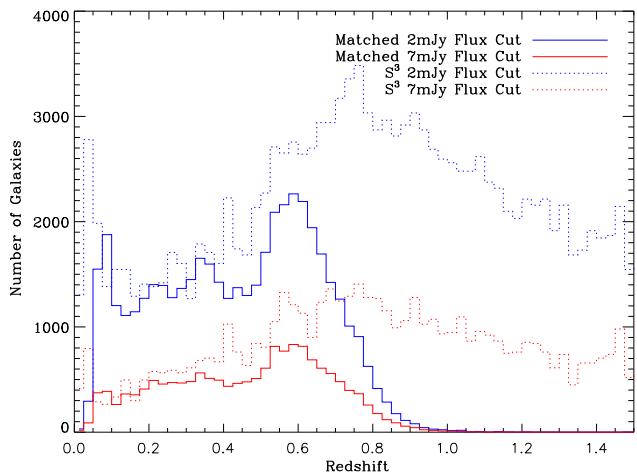}
  \caption{\small  The photometric redshift distributions of the 2 mJy
    (blue) \& 7 mJy (red) flux cuts of the FIRST sources that have
    been matched to the SDSS photometric survey (solid lines). The
    {\scubed} redshift distributions for the same cuts are shown as
    dashed lines. The distributions correspond a sky coverage of
    4613.43~${\rm deg}^2$ and the {\scubed} sample has been scaled
    accordingly to reflect this. } 
  \label{fig:dndz}
 \end{center}
\end{figure}

\subsubsection{Matching to the SDSS galaxies}
\label{sec:Matching_SDSS}
To match our FIRST sample to their optical counterpart we used data from the SDSS-DR7
 \citep[see e.g.][for a description of the seventh data
release]{Abazajian:2009ApJS..182..543A}.  In broad terms, the SDSS has mapped a quarter of the entire sky with
unprecedented accuracy using multi-band photometry ($u, g, r, i$ and
$z$) from the 2.5-meter telescope on Apache Point to a limiting
magnitude of $r<22.2$. The second phase of the project is now complete and is ideally
suited to our studies as it is fully contained within the FIRST survey area.

\AF{The number density of SDSS-DR7
photometric sources is orders of magnitudes greater than the density of
2 mJy FIRST sources. The average size of SDSS galaxies is between 2"
and 5s. To avoid erroneous matches we have
chosen a relatively conservative matching radius of 2"  
to match our FIRST sample to the SDSS-DR7 photometric 
catalogue. \SPnew{To ensure accurate matches we only consider objects classified
by the SDSS pipeline as a galaxy, requiring that they are successfully deblended to obtain precise positions, and have reliable photometric measurements in all 5 SDSS filters. 
Redshifts for the matched SDSS galaxies are taken from \citep{Oyaizu:2008}. Specifically, we use the photometric redshift estimated from a Neural Network method inferred from the 4 galaxy colours and 3 concentration indices. This estimate is recommended for faint ($r>20$) galaxies, which dominate the matched galaxy sample. Finally, we apply a minimum redshift cut of $z>0.01$ to remove contamination from misidentified stars. }
} 

\AF{It should also be noted that we are likely to miss some of the optical
identifications of fairly nearby multi-component radio sources as the
collapsed source position may not give the position of the optical
counterpart accurately enough. These sources are included in the
redshift distribution of the simulations (but not in the matched
redshift distribution) and thus will be included correctly in the
unmatched redshift distribution. Our method for probing the average
bias of the unmatched sample is thus still valid, but this effect could
make the interpretation of the average bias more complicated.}

\AF{Thus, for our central analysis we use four samples: 
45,883 FIRST matched galaxies, 173,177 FIRST unmatched galaxies with fluxes greater
than 2 mJy, and similarly 15,842 matched (77,360 unmatched) galaxies
with fluxes greater than 7 mJy. We probe the evolution of the bias in 
the matched sample by considering three
redshift bins corresponding to $0.01\le z<0.31$, $0.31\le z<0.56$ and
$z>0.56$, which were  chosen such that each bin contains approximately the same number of galaxies
(see Table~\ref{table:matching} for a summary).} 
\begin{table}
\footnotesize
  \begin{center}
    \tabcolsep 2pt
  \caption[Table detailing the matched data with  7 mJy and 2 mJy flux
  cuts.]{\MS{Details of of the number of sources passing each stage of our analysis for the matched and unmatched data} with a  7 mJy and 2 mJy flux
  cut.}
\label{table:matching}
      \begin{tabular}{llcllcllc}

      \\
      \hline
      \hline
      ~~~Matched/unmatched samples & ~~~~7 mJy cut &~~~~ 2 mJy cut &
             \\
      \hline
      \hline
\\
~~~Total number of sources		&~~~~~ 93,202		&~~~~ 219,060\\
~~~SDSS matched			&~~~~~ 15,842		&~~~~45,883\\
~~~SDSS unmatched		&~~~~~ 77,360		&~~~~173,177\\
~~~Redshift cuts of matched: 	&\\
~~~   \ \ $0.00\le z<0.31$		&~~~~~ 4334		&~~~~14,488\\
~~~  \ \ $0.31\le z<0.56$		&~~~~~ 5491		&~~~~15,533\\
~~~  \ \ $z>0.56$			&~~~~~ 6017		&~~~~15,862
\\
\\
      \hline
    \end{tabular}
  \end{center}
\end{table}
\begin{figure*}
  \begin{center}
    \includegraphics[width=0.49\textwidth]{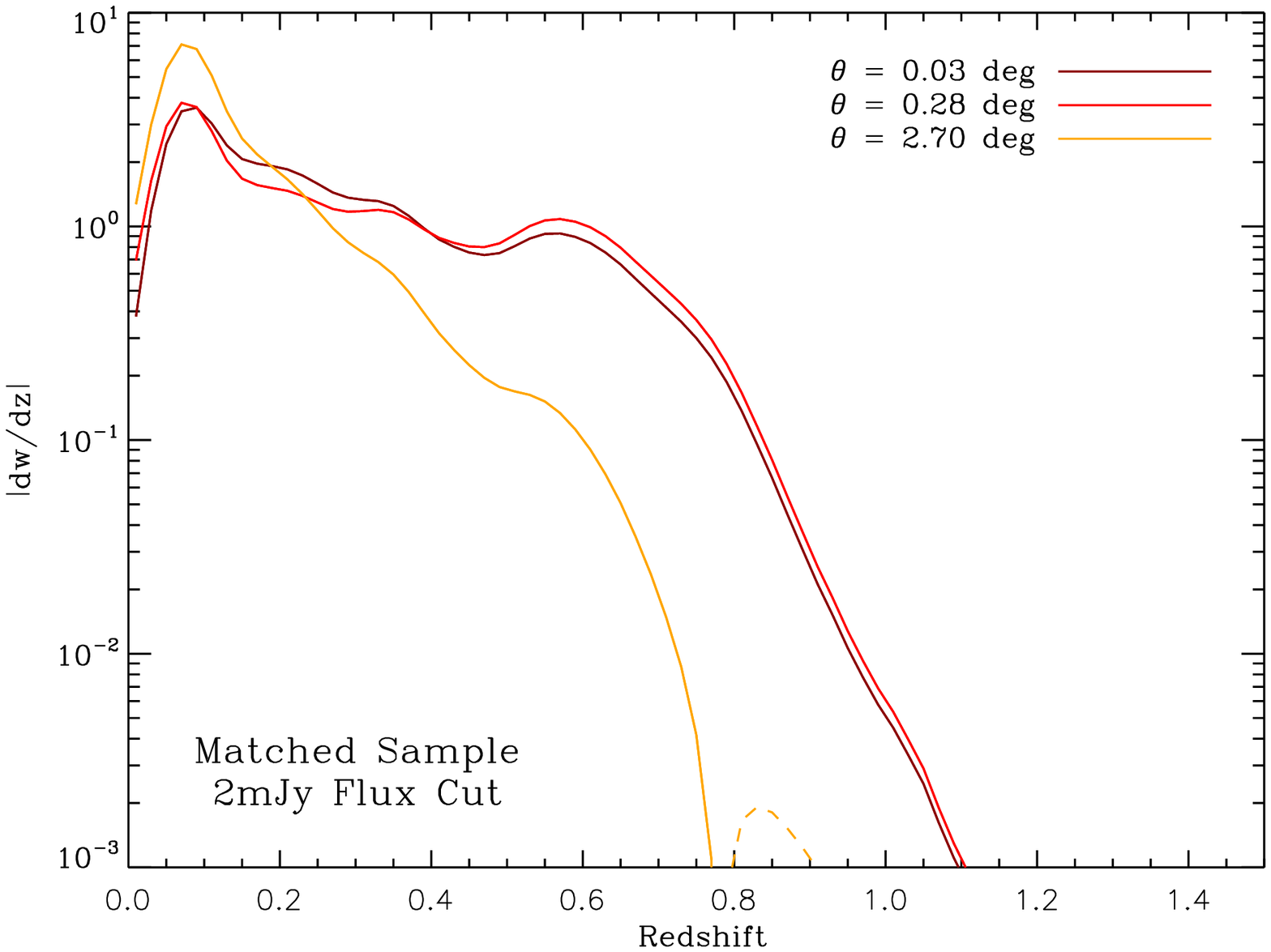} \hfill
    \includegraphics[width=0.49\textwidth]{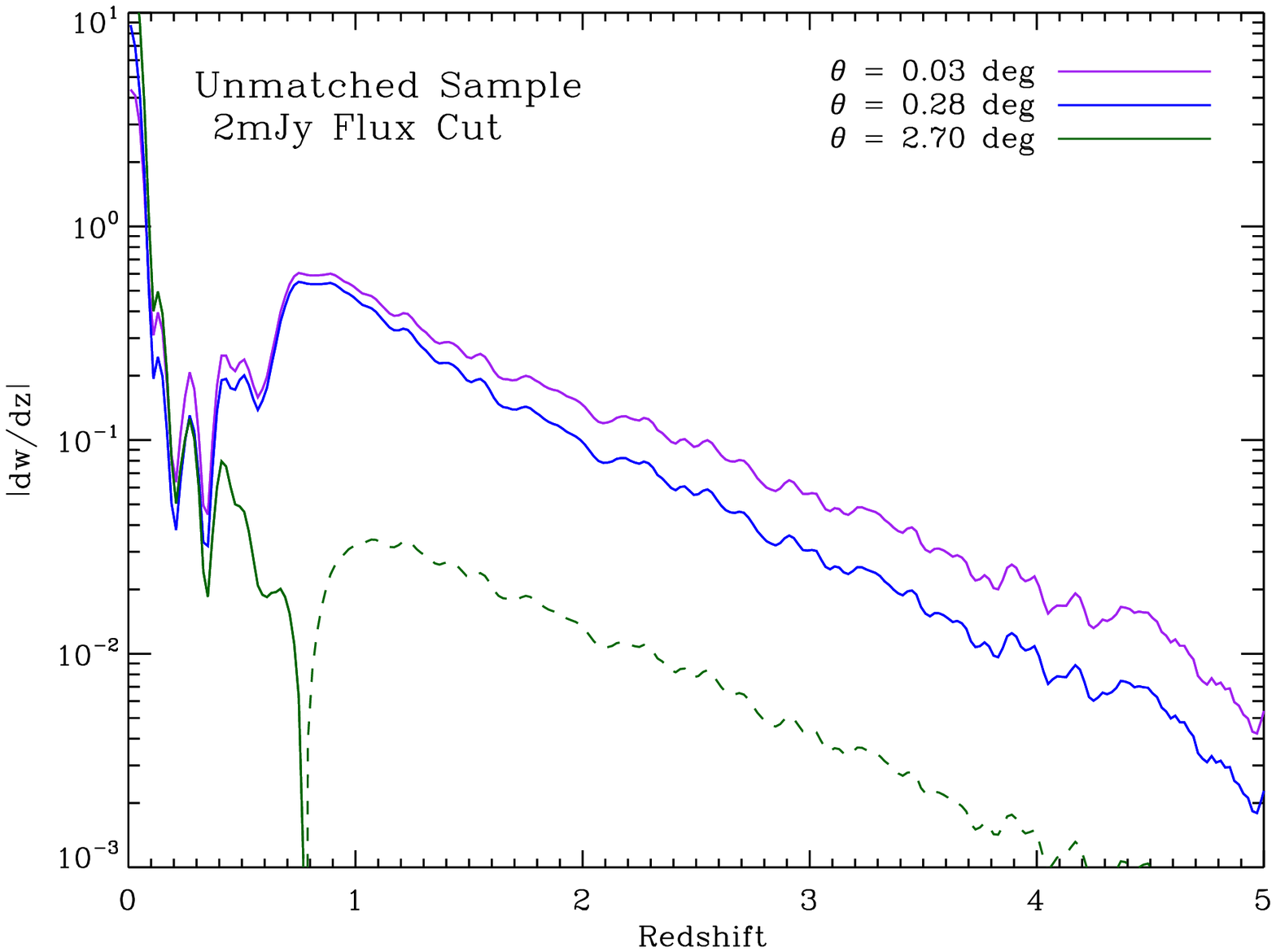} \\
    \vspace{5mm}
    \includegraphics[width=0.5\textwidth]{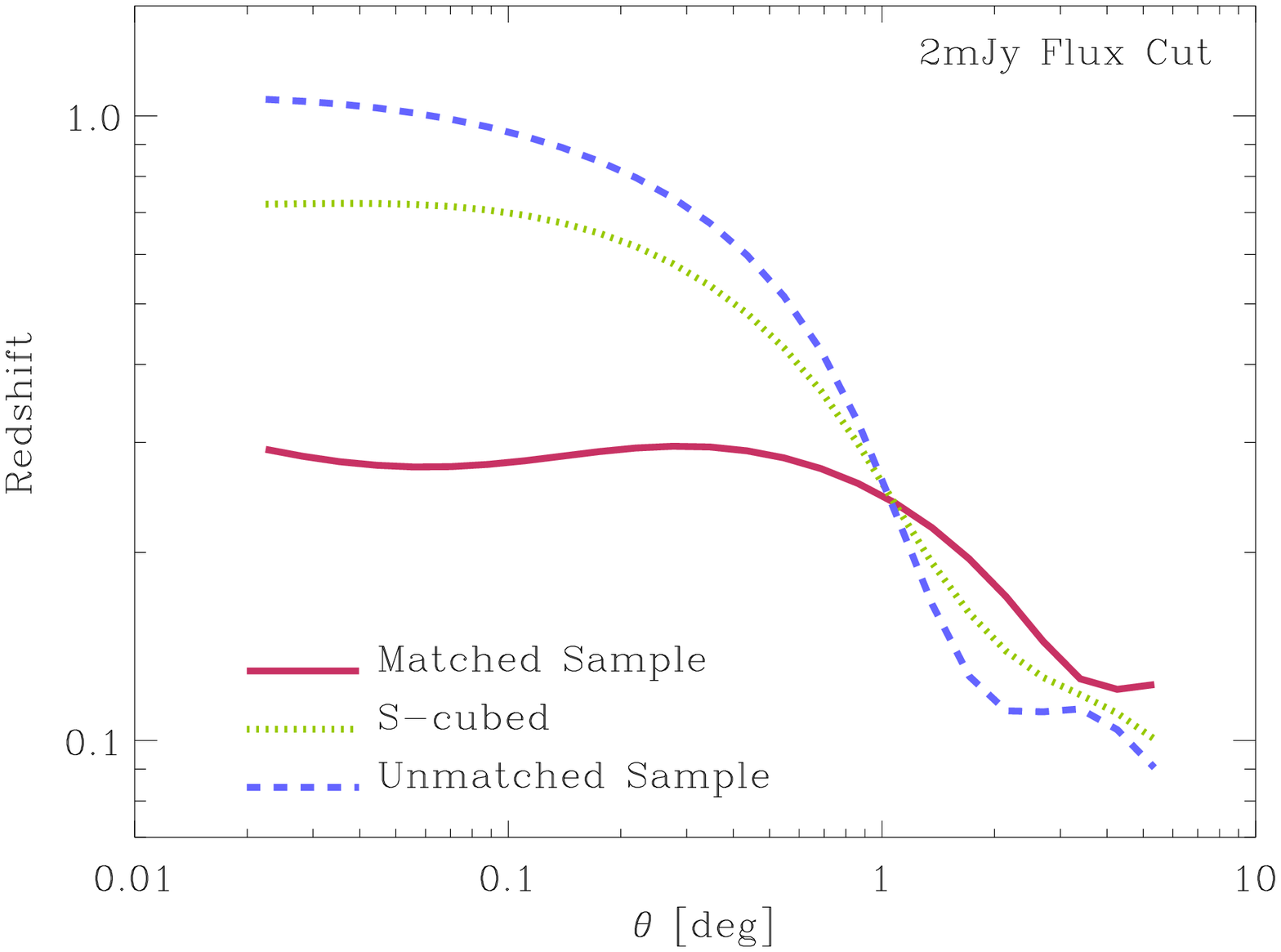}
    \caption{\small  \AF{The top panels show contribution to the
      ACF ($d\omega_{\rm DM}/dz|_{\theta}$) at three different angles
      ($0.03^\circ$, $0.28^\circ$, $2.70^\circ$) as a function of redshift. 
      Results are shown based on the redshift distributions of the
      matched and unmatched 2~mJy samples. The lower panel presents the
      average redshift, $\tilde{z}(\theta)$, which is probed at given angle
      for the three different samples indicated.}} 
    \label{fig:dm_pred}
  \end{center}
\end{figure*}

\subsection{Redshift distribution comparison}
We now compare our matched redshift distributions to that of the \AF{publicly available 
semi-empirical simulation of extragalactic radio continuum sources
  (\scubed) by \cite{Wilman2008MNRAS.388.1335W} which is part of the  
SKA Simulated Skies (S$^3$) project.
The {\scubed} covers} a sky area of $20\times20$ deg$^2$, out to a
cosmological redshift of $z = 20$. The simulated sources were drawn
from observed (or extrapolated) luminosity functions and grafted onto
an underlying dark matter density field with biases which reflect
their measured large-scale clustering. For each source, \MS{which include FRII galaxies, FRI galaxies, radio-quiet quasars,
starburst galaxies and star forming galaxies}, the database
gives the radio fluxes at observer frequencies of 151 MHz, 610 MHz,
1.4 GHz, 4.86 GHz and 18 GHz, down to flux density limits of 10
nJy. \MS{A prescription for clustering that captures the clustering pattern on
large scales (larger than those where non-linear evolution of density
fluctuations becomes important) was used}. The simulations can be used to
predict the redshift distribution of sources as a function of the flux
cutoff of surveys. 

Figure~\ref{fig:dndz} shows the redshift distributions for our matched
samples (solid lines) at the 7~mJy (red) and 2~mJy cuts (blue),
compared to the {\scubed} simulation (dotted line) for the same flux
cuts. In general we find agreement between the observed matched and
simulated redshift distributions up to $z\sim0.5$. However, we do note
that the prominent low redshift spike observed in the {\scubed} data
at $z\sim0.04$ does not appear in our matched sample. 
\subsection{Clustering analysis}
There are three different estimators that are used in the determination of
two-point correlation function as originally developed by
\cite{Davis:1982}, \cite{Hamilton:1993ApJ417} and
\cite{Landy-Szalay-93}. For this work, we apply the
\cite{Landy-Szalay-93} estimator, as it reduces errors caused by edges
of catalogues and sub-samples during error calculation. This estimator
can be written in the form: 
\begin{equation}
\label{equ:estimator}
\omega(\theta)=
\frac{DD(\theta)-2DR(\theta)+RR(\theta)}{RR(\theta)}\ ,
\end{equation}
where $DD(\theta)$ counts the number of pairs in the observed data as
a function of angular scale. Similarly, $RR(\theta)$ counts the number
pairs for the random catalogue and $DR(\theta)$ is the number of cross
pairs between data and random catalogue. The integral constraint is negligable.

For our analysis we populated our random catalogue with 50 times the
number of sources contained in the data for the matched and unmatched
samples, and 100 times the data from the three redshift bins (cf.,
Table~\ref{table:matching}). The errors on $\omega$ were calculated
using jack-knife re-sampling \citep{Lupton-93}. In this approach the
data was split into $N=24$ \MS{bins in RA} and the correlation
function is recalculated repeatedly each time leaving out a different
bin. A set of $N$ values $\{ \omega_i,i=1,...,N \}$ for the
correlation function are obtained and the jack-knife error of the
mean, $\sigma_{\omega_{mean}}$, is calculated by 
\begin{equation}
\sigma_{\omega_{mean}} = \sqrt{(N-1)\sum^N_{i= 1}(\omega_i-\omega)^2/N}\ .
\end{equation}

Each of the 24 bins can be
considered to be fairly independent due to the physical separation at the
redshift probed. 

In order to avoid problems associated
  with the over-cleaning of sidelobes, which effects the correlation
  function at $\theta \sim 0.2^\circ$, and any potential problems
  associated with collapsing multi-component sources, we only examine
  clustering at angles $\theta > 0.4^\circ$. \CCnew{We are also concerned that measurements at angles larger than $\theta>1^\circ$ may be unreliable (see section 3). }
\subsection{Clustering predictions from CDM}
\begin{figure*}
 \begin{center}
 \includegraphics[width=0.49\textwidth]{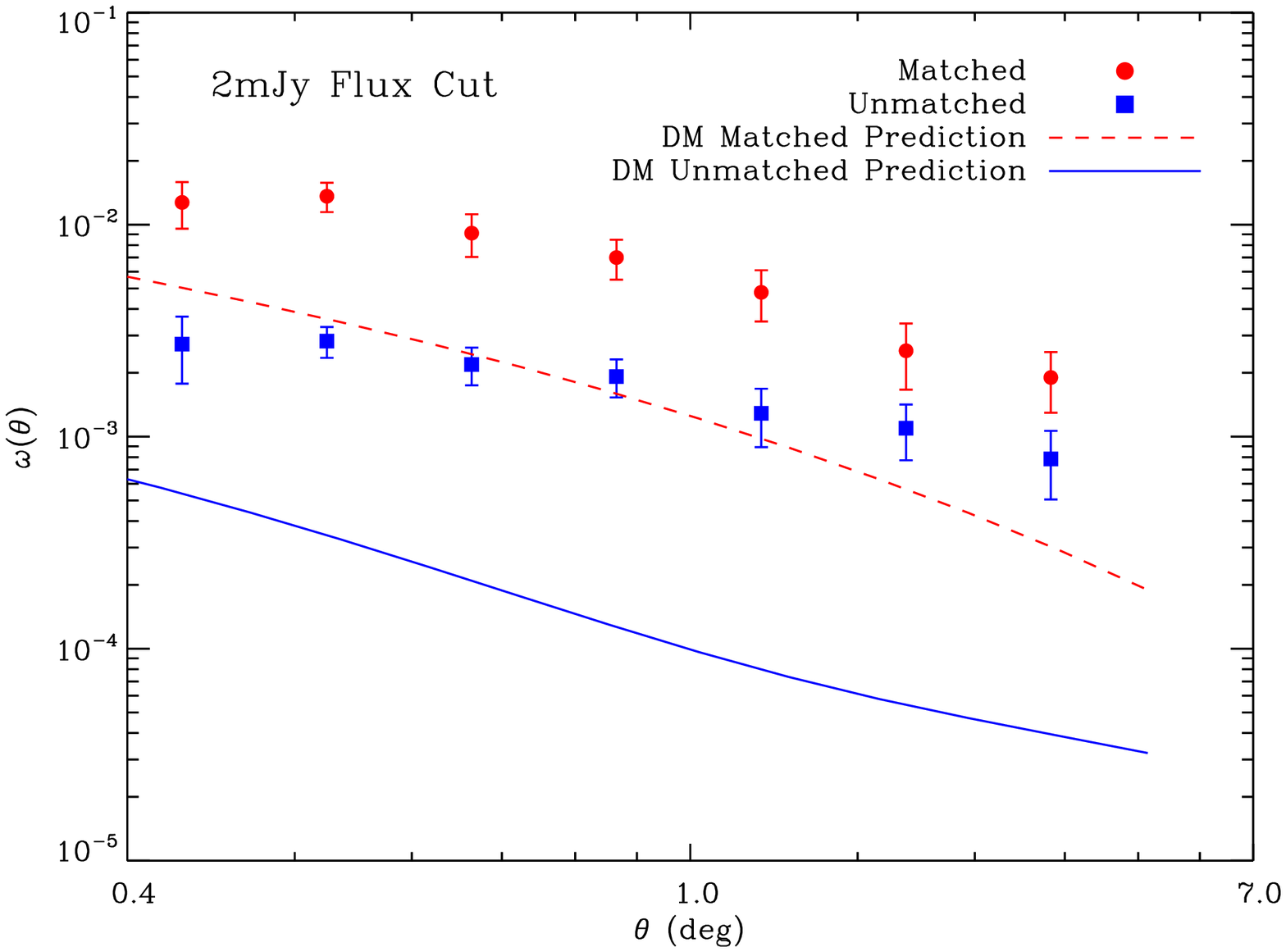}
 \includegraphics[width=0.49\textwidth]{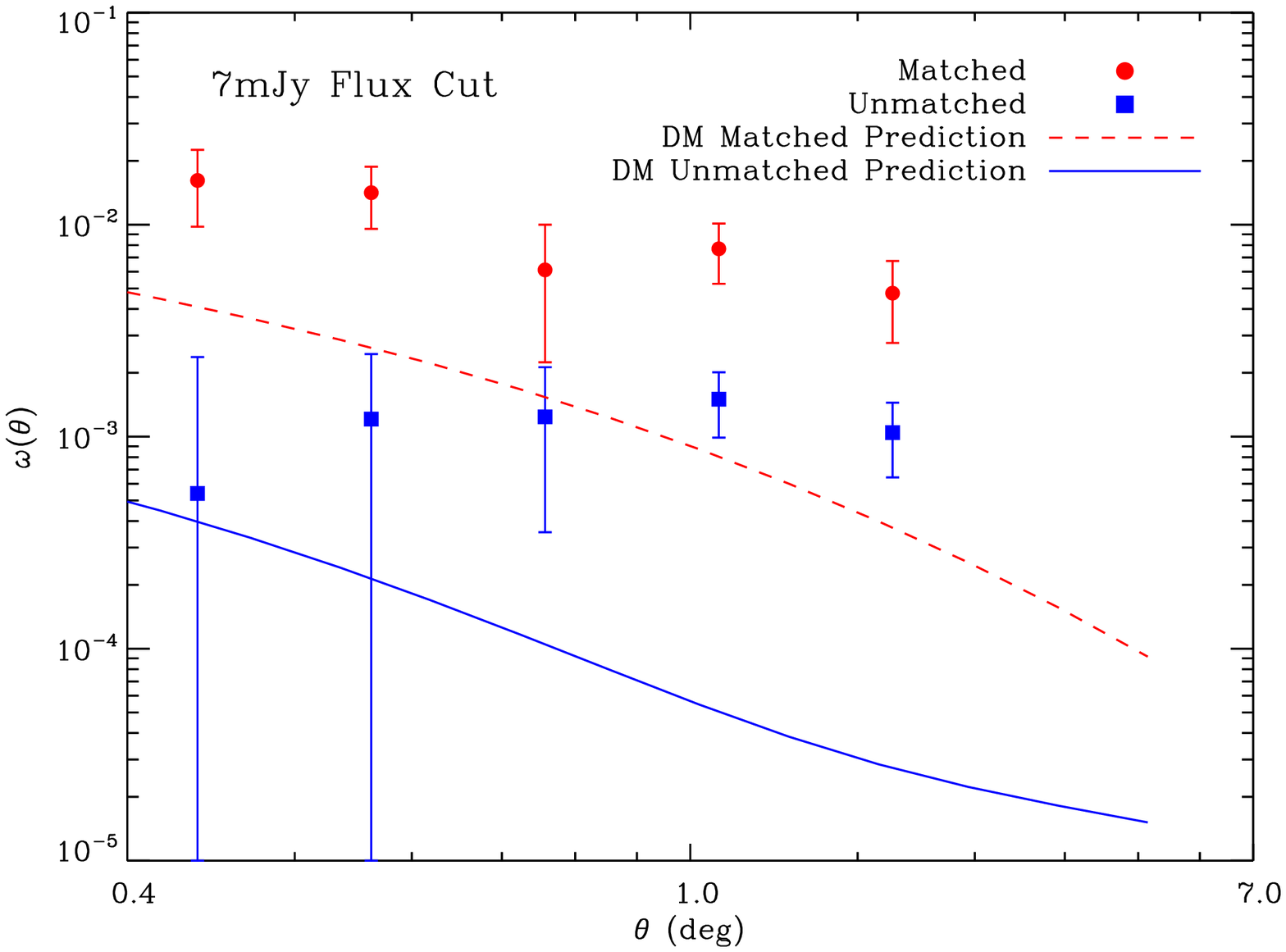}
 \includegraphics[width=0.49\textwidth]{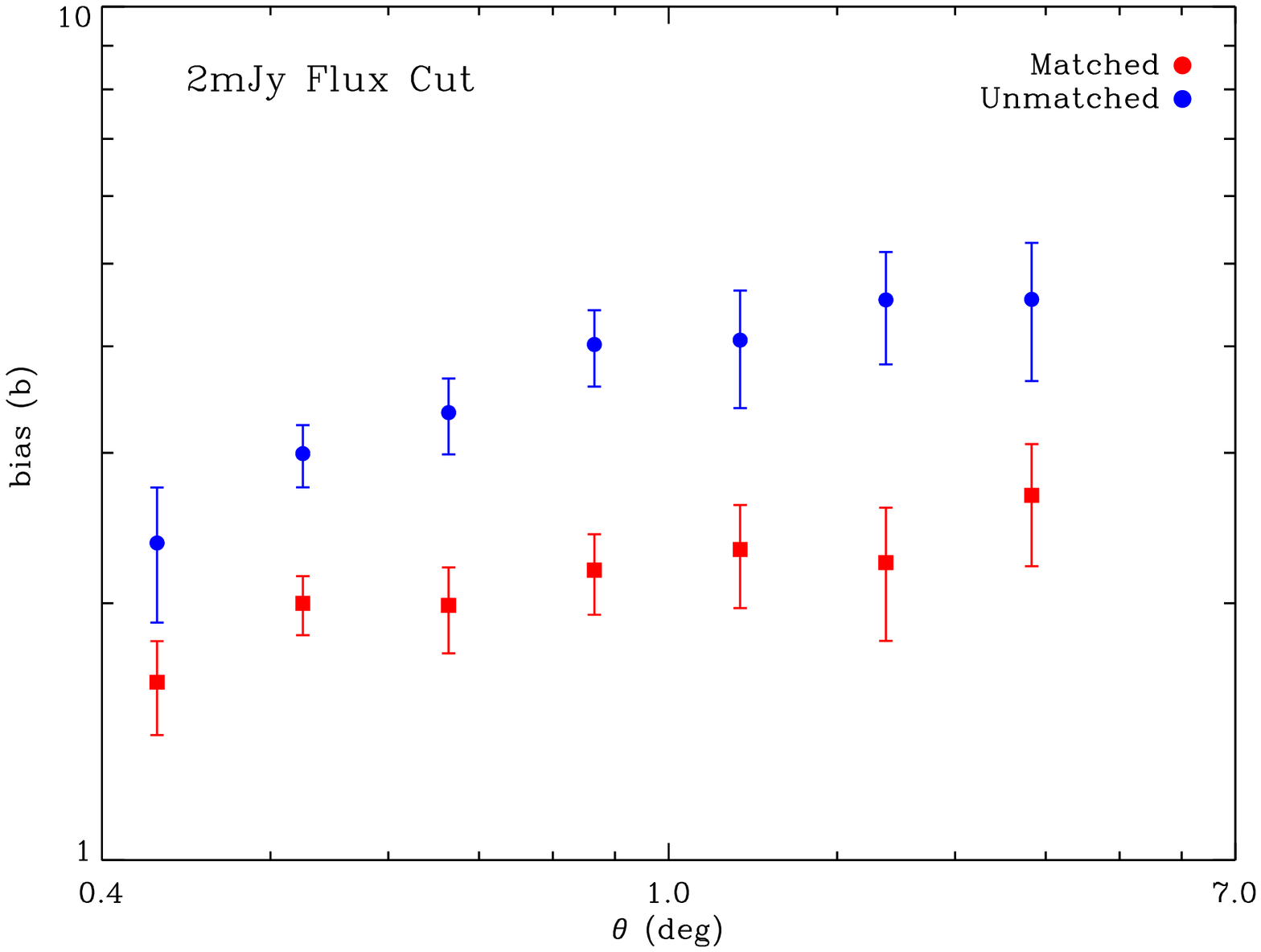}
  \caption{\small The two-point angular correlation function for the 2 mJy
(left panel) and 7 mJy (right panel) matched and unmatched samples. In both
panels the matched samples are indicated by blue points and the unmatched sample
by the red points. The corresponding dark matter (DM) predictions are shown
respectively by the dashed and solid lines. \CCnew{In the lower panel, we show the bias calculated for the 2 mJy
      flux cut of the matched (red) and the unmatched (blue) samples.} } 

\label{fig:2pcf}
 \end{center}
 \end{figure*}

To determine the bias of the radio population we compare their ACF with the corresponding dark matter correlation
function. If  $q(z)$ is the normalised redshift distribution of a
population of radio galaxies, the dark matter ACF can then be predicted from the non linear dark
matter power spectrum ($P_{\rm DM}$) via Limber's equation. For
spatially flat cosmologies one derives the following expression.
\begin{equation}
\omega_{\rm DM}(\theta) =
\int dr\ q^2(r )
\int \frac{dk}{2\pi} k\ P_{\rm DM}(k,z)\ J_0[r(z)\theta k]
\end{equation}
where $q(r)dr=q(z)dz$, $J_0(x)$ is the zeroth-order Bessel function of the first kind
and $r (z)$ is the radial comoving distance. Here we adopt the fitting
function for the non-linear CDM power spectrum by \cite{Peacock-Dodds-96}
using cosmological parameters given in \cite{Komatsu:2009}.

\CC{The linear bias, $b$, can be written} 
\begin{equation}
P_{lum}(k,z)=b^2(z,k)P_{\rm DM}(k,z)
\end{equation}
\CC{where $P_{lum}$ is the power spectrum of luminous tracers of the dark matter. Here, we measure a bias parameter, $b_\theta$, in the angular clustering signal which samples $b(k,z)$ for radio sources in FIRST:}
\begin{equation}\label{equ:bias}
b_\theta=\sqrt{\frac{\omega_{\rm gal}}{\omega_{\rm DM}}} .
\end{equation}

The derivative $d\omega_{\rm DM}/dz|_{\theta}$ at a given redshift $z$
reveals the contribution of that redshift slice to the overall ACF at the
angle $\theta$. \AF{The upper panels of Figure~\ref{fig:dm_pred} show 
  $d\omega_{\rm DM}/dz|_{\theta}$ as a function of redshift for the matched and unmatched
  samples (left and right panel, respectively) at three different
  angles ($0.03^\circ$, $0.28^\circ$ and $2.70^\circ$).} Based on that
one can determine the average redshift, $\tilde{z}(\theta)$,
which is probed at an angle $\theta$ for a given $q(z)$ by,
\begin{equation}
\tilde{z}(\theta) = 
\frac{\int z\ d\omega_{\rm DM}/dz|_{\theta}\ dz}{\int d\ \omega_{\rm DM}/dz|_{\theta}\ dz} \ .
\end{equation}
\AF{The lower panel of Figure~\ref{fig:dm_pred}
shows $\tilde{z}(\theta)$ based on the redshift distributions of the
SDSS matched and unmatched samples and the overall set of {\scubed} sources. 
For small angels, $\sim 0.1^\circ$ the (un)matched sample probes redshifts
of $z\sim0.3 (1.0)$. For angles above $1^\circ$ the average redshift
probed is below $0.25$ irrespective of which sample is considered.}    
\section{Results}
\label{sec:results}
\subsection{The angular two-point correlation function (ACF)}
%

\begin{figure*}
  \begin{center}
    \includegraphics[width=0.49\textwidth]{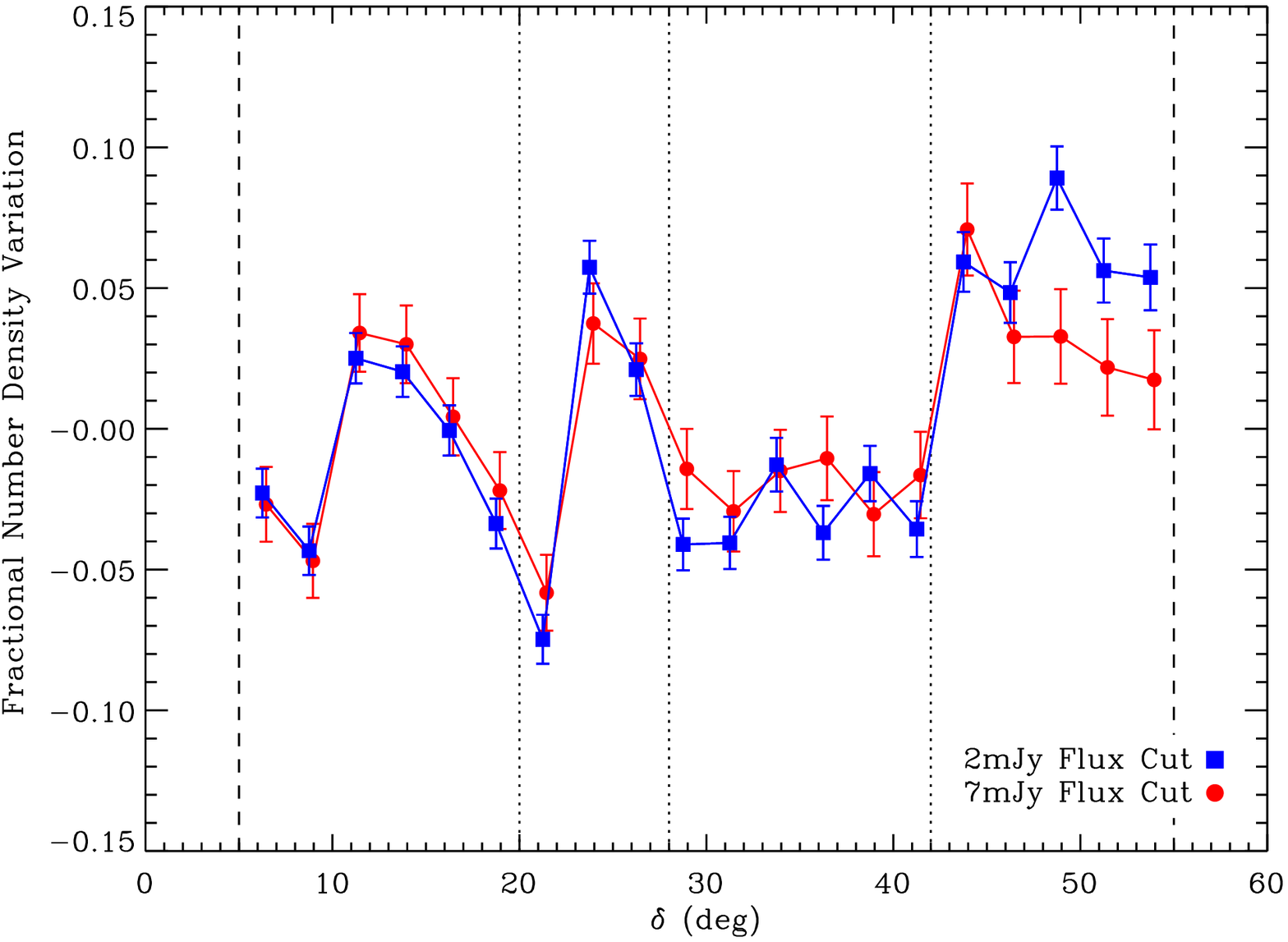}
    \includegraphics[width=0.49\textwidth]{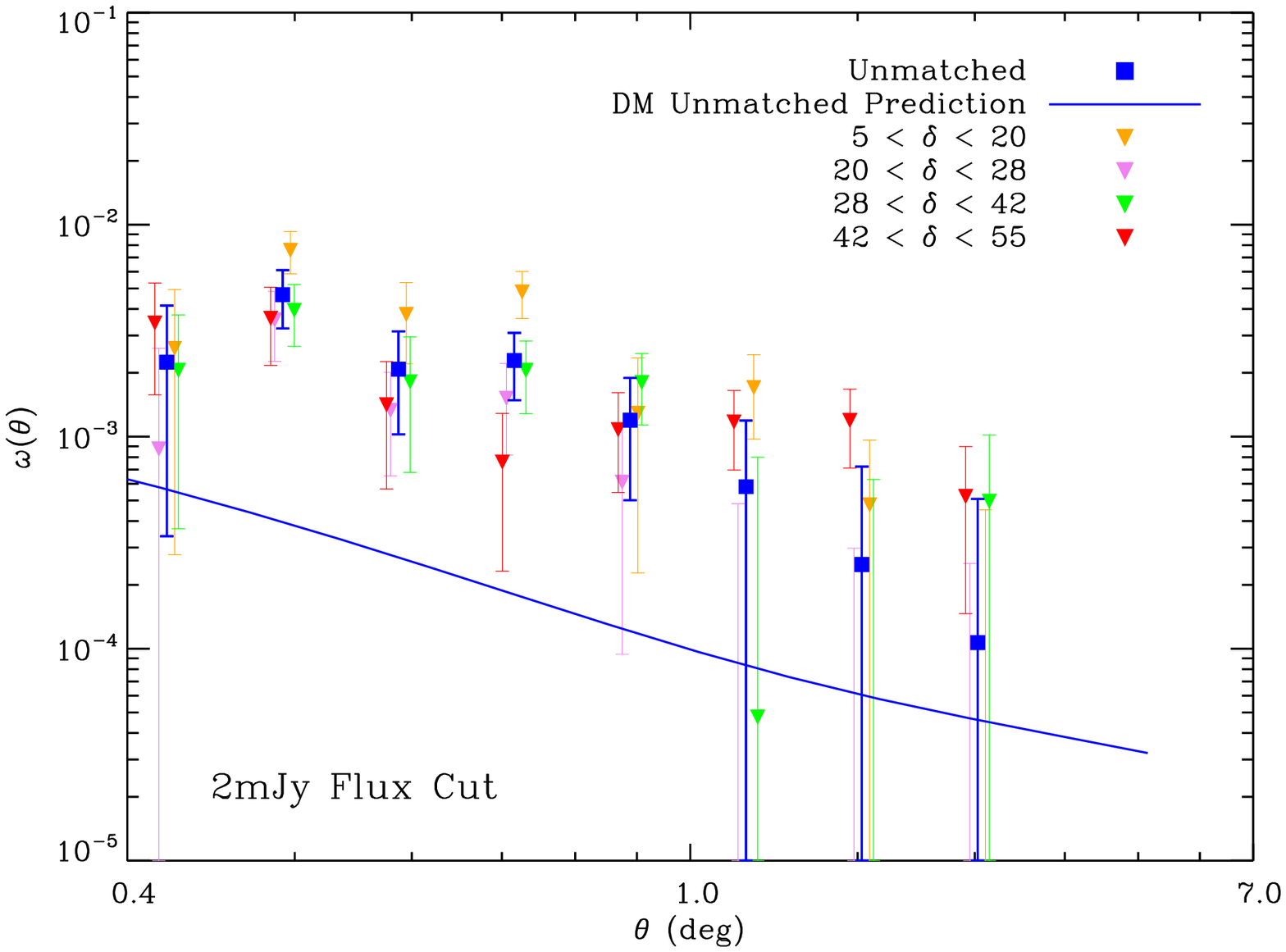}    \caption{\small In the left panel we plot the fractional number density variation of the source density, with vertical lines indicating the declination strips used in the right panel . In the right panel we plot the ACF measured in four different declination strips roughly corresponding to different observing epochs: ($5^\circ - 20^\circ, 20^\circ - 28^\circ, 28^\circ - 42^\circ$ and $42^\circ - 55^\circ$) }
    \label{fig:decstrips}
  \end{center}
\end{figure*}

Figure~\ref{fig:2pcf} shows the ACF for the 2mJy
(left) and 7mJy (right) matched (red circles) and unmatched (blue
squares) samples. In each panel the dark matter (DM) predictions are
shown as dashed and solid lines, respectively.
The bias (Equ.~\ref{equ:bias}) is  computed from the ratio between
the data and predicted dark matter correlation functions and is shown
as a function of angle in 
the lower panel for the 2mJy sample. 

For the 2mJy cut, we see that the matched sample is more clustered (in
angular projection) than the unmatched sample. This is expected \AF{since 
the matched sample occupies lower redshift ranges (cf.,
Fig.~\ref{fig:dndz}), thus a given angle
corresponds to smaller physical scales where there is more
clustering. The ACF for the full 1~mJy sample found 
by \cite{Cress-96} lies between our matched and unmatched curves}.
The
amount of clustering measured in both the matched and unmatched
samples at angles greater than $1^\circ$ is difficult to explain when
one considers the results in Figure~\ref{fig:dm_pred}. On these
scales, one would expect to probe $z\sim 0.1 $ where the sample
contains many fainter star-forming galaxies with a bias similar to
normal galaxies i.e. \CC{$b_\theta\sim 1$}. Instead, we see a bias \CC{$b_\theta> 4$} for the
unmatched and values \CC{$b_\theta>2$} for the matched sample. 

To explore the possibility that large angle fluctuations are due to systematic 
variations in source density associated with different observing epochs, 
we plot \seannew{fractional number density variation as a function of declination in the left panel of Figure~\ref{fig:decstrips} and note some fairly large changes in the 
fractional number density. }
To investigate the impact of this on the correlation function measurements, we divide the FIRST sources into declination strips roughly associated 
with different observing epochs and calculate the ACF in each strip. 
The results shown in the right panel of Figure~\ref{fig:decstrips} indicate that, beyond $1^\circ$, the results in the different declination strips start to differ. 
This suggests that systematics might have a significant effect on larger scales. However, we note that on smaller scales the measurements are consistent with each other, indicating that these scales are free of systematics related to this effect.  We discuss other possible explanations for the excess large scale-power seen in the full sample in section 3.2  .

The right panel of Figure~\ref{fig:2pcf} shows the ACF for the 7mJy sample, which should be completely dominated by AGN \CCnew{\citep{Best-03}}. The clustering of the matched
sample is consistent with that of the 2mJy matched sample. In the unmatched sample, the bias is higher at large
angles. The low measurements at smaller angles may indicate that the
sidelobe over-cleaning problem is more pronounced for brighter
sources.

\begin{figure*}
  \begin{center}
    \includegraphics[width=0.49\textwidth]{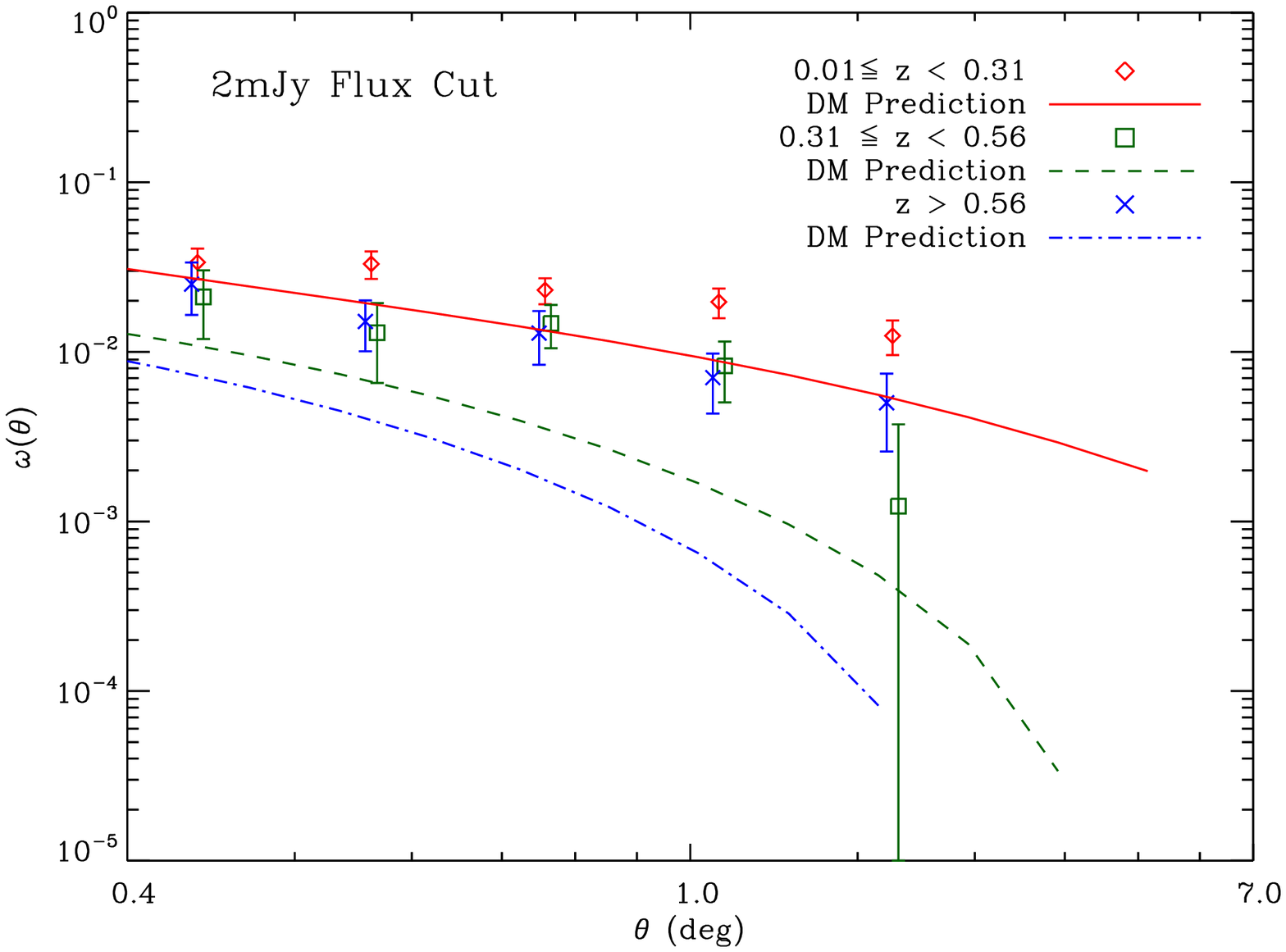} 
    \includegraphics[width=0.49\textwidth]{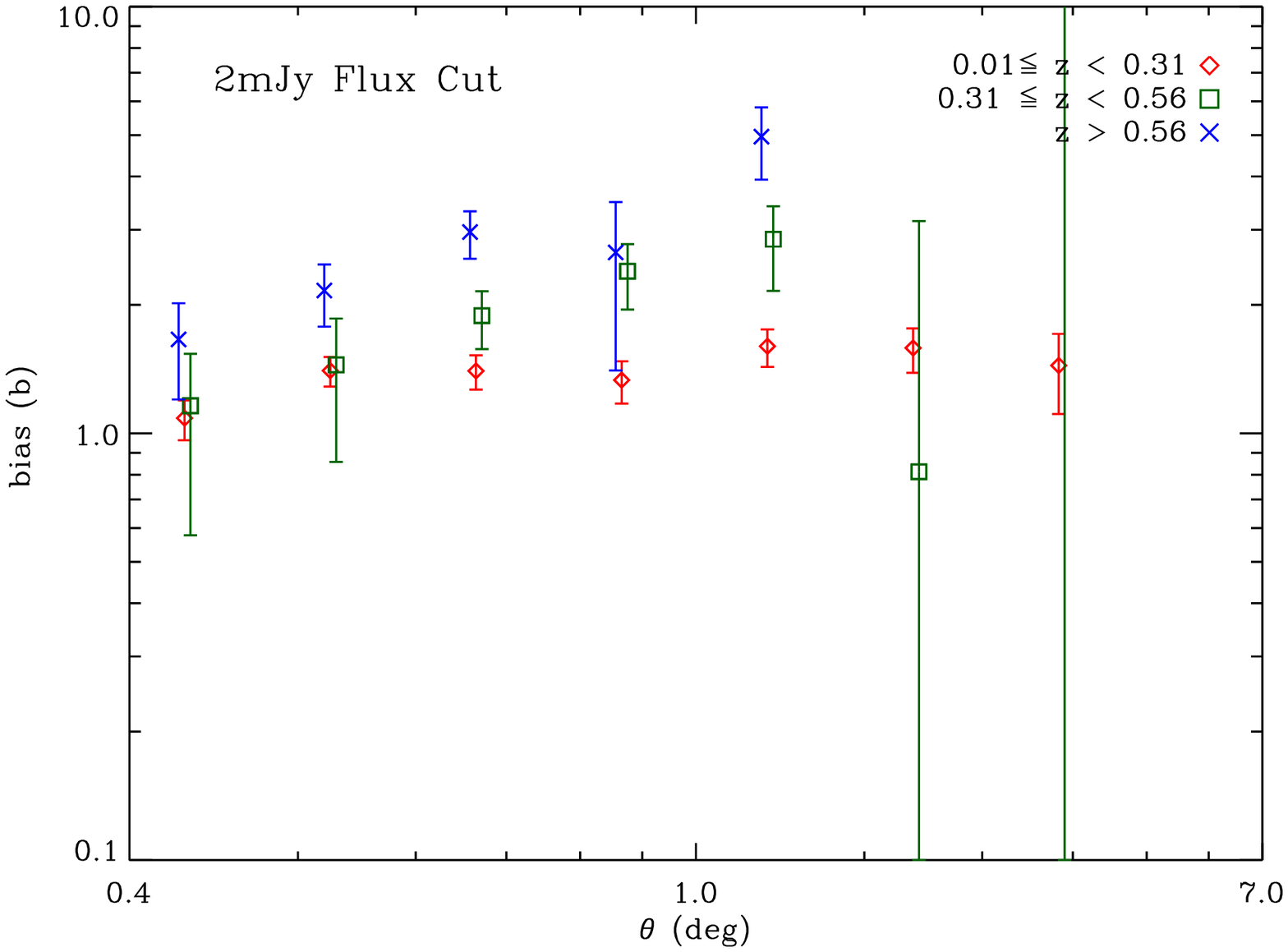}
    \caption{\small The left panel shows the ACF for the 2 mJy matched sample split
      into three redshift slices maintaining approximately the same
      number of objects in each slice. The three slices correspond to:
      $0.01\le z< 0.31$, (shown as red diamonds) $0.31\le z < 0.56$
      (green squares) and $z > 0.56$ (blue crosses). For each slice we
      have plotted the corresponding dark matter (DM) prediction. The right
      panel shows the evolution of bias for the three redshift slices. } 
    \label{fig:2pcf_zslice}
  \end{center}
\end{figure*}

To help interpret the matched ACF, we split the 2 mJy matched sample
into three redshift slices, keeping the number of sources in each
slice approximately constant. In the left panel of Figure~\ref{fig:2pcf_zslice}, we plot
the ACF for each of the redshift slices and in the right-hand panel
we plot the bias calculated as a function
of angle for each slice. One sees that the bias for the lowest
redshift slice is fairly close to \CC{$b_\theta\sim1$}, as one would expect for a
population dominated by fairly ordinary star-forming galaxies. Sources
in the highest redshift bin are much more biased, as one would expect
for a population dominated by AGN that trace large halo masses in the
universe. The important point to note is that according to 
Figure~\ref{fig:dm_pred} the average redshift
probed for the matched sample at larger angles is about $z\sim0.12$,
but we see a large bias for the matched sample, left panel in
Figure~\ref{fig:2pcf}, at these angles and this \MS{can} be
attributed to the more highly biased population at $z>0.31$. 

Given that our main aim in this work is to constrain the bias toward
high redshifts, we choose an angle of $0.66^\circ$ to determine the
clustering behaviour of the high redshift radio sources. According to
Figure~\ref{fig:dm_pred} this choice allows us to probe bias at
$z\sim0.7$ . 
\MS{We find that the unmatched sources are more biased than the matched sample 
(at $3.3\sigma$), with a value of \CC{$b_\theta=3.0 \pm 0.25$}, compared to $2.0\pm 0.16$ for the 
matched sample at a mean redshift of $z\sim 0.7$.} 
\begin{table}
\footnotesize
  \begin{center}
    \tabcolsep 2pt
  \caption{Bias results  measured at an angle of 0.66$^\circ$  for the matched,
unmatched and the three
  redshift bins.}\label{table:bias}
      \begin{tabular}{llcll}
      \\
     	 \hline
           \hline
      ~~~Samples &~~~~ Bias (\it{$b_\theta$})
      \\
      	\hline
      	\hline
\\
~~~Matched			&~~~~2.0 $\pm0.16$\\
~~~Unmatched			&~~~~3.0 $\pm0.25$\\
~~~$0.01\le z<0.31$		&~~~~1.4 $\pm0.16$ \\
~~~$0.31\le z<0.56$		&~~~~1.5 $\pm0.50$ \\
~~~$z>0.56$			&~~~~2.2 $\pm0.35$ \\
\\

      \hline
    \end{tabular}
  \end{center}
\end{table}
\subsection{Excess power at large angles}
In this paper, our results are based on measurements at angles smaller than $1^\circ$ but it is interesting to consider explanations for the excess power at larger
angles in the $unmatched$ sample.
\begin{enumerate}
\item {Following the discussion for the
matched sample, we could reason that, a highly biased population at
high redshift could contribute significantly to the measurement at
$\theta> 1^\circ$, even though Figure~\ref{fig:dm_pred} indicates that
the average redshift probed on large angles is small. Bias of $b_\theta>4$,
however, is not seen even for fairly massive clusters and additional
contributors should be considered}

\item \CC{Systematics other than those discussed in section 3.1 could also contribute. The beam shown in \cite{Condon-98} for the NVSS survey does not go to zero at large angles, suggesting that bright sources could produce artefacts at large angles due to imperfect cleaning in VLA data. However, similar `excess power' is observed in the SUMSS radio survey
  which was carried out using a very different kind of
  telescope \citep{Blake-Mauch-Sadler-04}. Nevertheless, there is a possibility that radio surveys contain spurious sources which are correlated on large angles and this is a possible explanation for the excess power observed in clustering studies. }  
    
\item There is a low-redshift spike in the source counts, not included
  in the redshift distribution used for the dark matter
  predictions. This would push up the clustering amplitude on all
  angular scales, but particularly on the larger scales. However, the
  similar behaviour of the 2 and 7mJy indicate that the excess power
  is not due to faint, low-z star-forming galaxies. Also, the results
  of \citealt{Magliocchetti-04} and \citealt{Mauch-Sadler-07} appear
  to rule out this explanation. \AF{The {\scubed} redshift distribution
  which we use here is designed to fit these observations. A
  hypothetical low-redshift population which would have been missed in
  these studies would need to have $K>12.75$ and $B>19.45$, making
  such a low-z obscured population an unlikely explanation for much of
  the excess power in the unmatched sample.}  
\item Our matching technique is likely to result in some low-z
  multi-component radio sources being missed in our matched sample
  and one would expect these sources to be more biased than ordinary
  galaxies. This could boost the amplitude of clustering on large
  scales.  
\item  Finally, there is the possibility that non-gaussian initial
  conditions could generate more clustering on large scales than in
  the standard model as suggested by \cite{Xia-10}.
\end{enumerate}

\CCnew{Further work is clearly needed to understand the excess power in the clustering signal on large angles.}
\subsection{Consistency checks}
We carried out a number of tests to check the robustness of our
results. In the first test, we changed the matching radius to 1" to decrease the number of false identifications. This did not
impact that ACF or the average redshift distribution of the unmatched
sample, indicating that the bias measurement at $z\sim 0.7$ is not
sensitive to the choice of matching radius. In the second test, we
used the matched sample of \cite{Best-03} rather than our own
matching. This sample was carefully constructed using both NVSS and
FIRST and used visual identification rather than an automated
``collapse and match'' approach. Results were consistent with our
matched sample, given that their sample probes a somewhat different
redshift range to ours.  In the third test, we considered the impact of our
choice of sidelobe probability cut. We calculated the ACF for several
different sidelobe probability samples and found that all samples behaved
similarly at large angles. \MS{Finally, we investigated the sensitivity of our results to the photometric redshift estimates, by using  different
SDSS photometric redshift catalogues. We found that the results were robust to the choice of catalogue.}
.
\section{Conclusions}
\label{sec:con}
\MS{We have introduced a method for measuring the bias at high redshift for a sample of radio continuum sources lacking redshift information. }
By matching radio sources from  the FIRST survey data to their optical
counterparts in the SDSS survey, we extracted a subsample of  unmatched
objects. We  then used the {\scubed} simulation to infer an  average
redshift distribution for all FIRST sources and estimate the
redshift distribution of unmatched sources by subtracting the matched
distribution from the distribution of all sources.  

We have found that the surprisingly large clustering signal at large
angular scales present in the full FIRST  sample is also detected in
the unmatched sample considered here, and to some extent, in the
matched samples at high redshift.  \CCnew{We note that this could be due to systematic fluctuations in sensitivity in different observing epochs but also discuss a number of other possible explanations. } 
 Using clustering measurements at smaller angles, we estimate the bias of the unmatched FIRST sources
with flux densities over 2 mJy, at $z\sim0.7$, to be  $b_\theta=3.0\pm0.25$.

The analysis of cross-correlations with other data will be helpful in
interpreting these measurements better. These results can help constrain
models of radio source evolution and are important for using radio
surveys to constrain cosmological models.
\section{Acknowledgements}
The authors acknowledge the National Research Foundation
and the SKA (South Africa) funding bodies, as well as the South 
African Astronomical Observatory (SAAO) and the Institute of Cosmology 
and Gravitation (ICG)  where some of this work was carried out. The authors thank
the African Institute for Mathematical Sciences (AIMS) for hosting 
the initial meeting where this project was concevied. 
RJ and MS would like to thank Prina Patel, Cris Sabiu and Luis Teodoro for valuable discussions. 
\setcounter{equation}{0}
\renewcommand{\theequation}{A-\arabic{equation}}
\setcounter{section}{0}
\renewcommand{\thesection}{A-\arabic{section}}
\setcounter{figure}{0}
\renewcommand{\thefigure}{A-\arabic{figure}}

\label{lastpage}
\onecolumn
\end{document}